\documentclass[preprint,prb,aps,amsmath,amssymb,color,superscriptaddress]{revtex4}
\usepackage{stmaryrd}
\usepackage{amsmath}
\usepackage{amssymb}
\usepackage{graphicx}
\usepackage{dcolumn}
\usepackage{bm}
\usepackage{tabularx}
\usepackage{color}
\usepackage{booktabs}
\usepackage{multirow}
\begin{document}
\begin{titlepage}
\title{Giant enhancement of exciton radiative lifetime by ferroelectric polarization: The case of monolayer TiOCl$_2$}
\author{Hongwei Qu and Yuanchang Li}
\email{yuancli@bit.edu.cn}
\affiliation {Key Lab of advanced optoelectronic quantum architecture and measurement (MOE), and Advanced Research Institute of Multidisciplinary Science, Beijing Institute of Technology, Beijing 100081, China}
		
\date{\today}

\begin{abstract}
Exciton binding energy and lifetime are the two most important parameters controlling exciton dynamics, and the general consensus is that the larger the former the larger the latter. However our first-principles study of monolayer ferroelectric TiOCl$_2$ shows that this is not always the case. We find that ferroelectric polarization tends to weaken exciton binding but enhance exciton lifetime. This stems from the different effects of the induced built-in electric field and structural distortion by the spontaneous polarization: the former always destabilizes or even dissociates the exciton while the latter leads to a relaxation of the selection rule and activates excitons that are otherwise not optically active. Their combined effect leads to a halving of the exciton binding energy but a substantial increase in lifetime by 40 times. Our results deepen the understanding of the interaction of light with ferroelectric materials and provide new insights into the use of ferroelectricity to control exciton dynamics.
\end{abstract}

\maketitle
\draft
\vspace{2mm}
\end{titlepage}

\textbf{I. Introduction}

Excitons are electron-hole pairs bound by Coulomb interactions. They play a particularly important role in light emission, absorption and photovoltaic applications\cite{Klimov,Sambur}. In addition to this, on scales larger than their radii, excitons can be regarded as bosons and undergo Bose-Einstein condensation. Although it is theoretically proposed that excitons can form spontaneously and condense into excitonic insulators, decisive experimental evidence for this is still lacking so far\cite{Kogar,EI1967}. In most cases, excitons behave as an elementary excitation of condensed matter with finite lifetimes. While longer lifetimes are favourable for exciton condensation, longer lifetimes are not always better, for example in solar cells. A solar cell works on a system with a bandgap that absorbs sunlight to produce excitons, then these electron-hole pairs are dissociated by an electric field, and eventually electrons and holes move to the opposite electrodes where they are collected, completing the light-to-electric conversion\cite{Jarvist,Noel,Lee}. This process requires the exciton lifetime to be in a suitable range. Too long a lifetime means that a stronger electric field is required to achieve exciton dissociation; too short a lifetime means that electrons and holes recombine and radiate before reaching their respective collecting electrodes, hindering the photoelectric conversion. Controlling exciton lifetimes is therefore a subject of both scientific and practical interest.

Since the discovery of graphene in 2004, a wide variety of two-dimensional materials have been synthesized and explored. They exhibit exotic properties that are very different from their bulk counterparts. For example, MoS$_2$ evolves from an indirect-gap semiconductor to a direct-gap semiconductor when its number of layers is reduced to the limit of only a single layer\cite{Mak}. One of the authors of this work found that there is a linear scaling law between the exciton binding energy and the quasiparticle gap for two-dimensional materials\cite{Jiang2017}, which does not exist in three dimensions. The reduction in dimensionality leads to a weakening of the screening effect, so that two-dimensional materials typically have larger exciton binding energies, offering new opportunities for the long-sought excitonic insulators\cite{Dong2020,Dong2021,Jiang2018,Jiang2019,Jiang2020,Varsano}. The advent of two-dimensional materials has also revolutionized some of the traditional understanding of materials. One example is ferroelectric materials. Unlike polar structures without inversion centers, the presence of a spontaneous electric polarization in ferroelectrics reduces the energy of the centrosymmetric structure, leading to a spontaneous breaking of the inversion symmetry. The ferroelectric polarity can be switched by applying an electric field, resulting in a characteristic hysteresis loop. Since the depolarization field increases as the material becomes thinner, it was long thought that ferroelectricity would disappear below a critical thickness until stable ferroelectricity was found in atomic-thick SnTe down to a 1-unit cell limit\cite{Chang}. Ferroelectric switching was achieved even in two-dimensional metals\cite{Fei}, which was ruled out in conventional three-dimensional ones due to strong electronic screening against external electric fields.

Electric fields are commonly used to modulate the properties of low-dimensional excitons, such as binding energy, transition energy, linewidth and dissociation\cite{Chernikov,Raja,Massicotte,LiuJ,Ponomarev}. Electrical tuning of exciton binding energies up to a few hundred microvolts has been demonstrated in monolayer transition-metal dichalcogenides\cite{Chernikov}. It is clear that the electric field also affects the exciton lifetime, but very little is known about this. Ferroelectric materials have a built-in electric field due to the potential difference along the polarization direction generated by the dipole field. The presence of this built-in electric field makes the excitonic effect in ferroelectric materials fundamentally different from that in ordinary semiconductors. In particular, the observation of open-circuit voltages above the bandgap in BiFeO$_3$ films has stimulated enthusiastic research on ferroelectric materials for photovoltaic applications\cite{Yang,Zenkevich}. Whether the physical reason behind this abnormal ferroelectric photovoltaic effect is due to the bulk photovoltaic effect or to the electron-hole dissociation promoted by the built-in electric field at the ferroelectric domain is still under debate\cite{Seidel,Bhatnagar}. Understanding how the built-in electric field affects the excitonic properties naturally contributes to the understanding of the abnormal photovoltaic effect of ferroelectric materials.

In this work, we perform a first-principles study of the excitonic properties of monolayer ferroelectric TiOCl$_2$ using $GW$ plus Bethe-Salpeter equation (BSE) calculations. We compare the exciton binding energy and lifetime in its paraelectric and ferroelectric states, respectively, and find that spontaneous polarization leads to a halving of the binding energy but a 40-fold increase of the lifetime. By artificially adjusting the displacement of Ti, we are able to analyze the evolution of excitonic properties with the polarization strength. It turns out that the effect of ferroelectricity on excitons is reflected in two ways: i) the built-in electric field always weakens the binding of excitons or even dissociates them completely. ii) The induced structural distortion significantly enhances the atomic orbital hybridization and thus relaxes the selection rule. As a consequence, a new optical exciton with a smaller transition dipole is generated, presenting a longer lifetime. Our work provides theoretical insights into the use of ferroelectricity to control the exciton dynamics.

\vspace{0.3cm}
\textbf{II. Methodology and models}
\vspace{0.3cm}

All DFT calculations were performed within the Perdew-Burke-Ernzerhof (PBE)\cite{PBE} exchange correlation as implemented in the QUANTUM ESPRESSO package\cite{QE}. The energy cutoff for plane-wave basis is set to 60 Ry. A 15 \AA\ vacuum layer was used to avoid spurious interactions between adjacent layers. The single-shot $G_0W_0$ and BSE calculations\cite{GWBSE} were implemented using the YAMBO code\cite{Yambo1,Yambo2} with the Coulomb cutoff technique. After the convergence test, dielectric function matrix and quasiparticle band structure were calculated with an 18$\times$20$\times$1 Monkhorst-Pack $k$-point grid, 200 bands and 13 Ry cutoff. A more dense 24$\times$26$\times$1 grid was used for solving the BSE on top of the $G_0W_0$ band structure. Top four valence bands and bottom four conduction bands were used to build the BSE Hamiltonian.

Here, we concern about the exciton radiative lifetime $\tau$ at 0 K, which is derived within the Fermi's golden rule for a two-dimensional system\cite{Palummo1,Hsiao1,Hsiao2}
\begin{equation}
    \tau=\frac{A_{\mu c}\hbar^2c}{8\pi e^2E_S(0)\mu_S^2}
	\label{eq:0}
\end{equation}
where $A_{uc}$ is the area of the unit cell and $E{_S(0)}$ is the transition energy of the zero momentum exciton. ${\mu_S^2}$ is the square of the dipole matrix element\cite{Spataru}, which is defined as
\begin{equation}
	\mu_S^2 = \frac{{{\hbar ^2}}}{{{m^2}E_S(0)^2}}\;\frac{{\left| {\left\langle {G\left| {{p_{||}}} \right|{\Psi _S}(0)} \right\rangle } \right|}^2}{{{N_k}}}.
	\label{eq:9}
\end{equation}
Here $N_k$ is the number of $k$-points used, $m$ is the electron mass, and ${\left| {\left\langle {G\left| {{p_{||}}} \right|{\Psi _S}(0)} \right\rangle } \right|}$ is the exciton transition dipole\cite{Hsiao1,Hsiao2}. All these quantities can be directly deduced from the first principles calculations.

\vspace{0.3cm}
\textbf{III. Results and discussion}
\vspace{0.3cm}

\begin{figure}[htb]
\includegraphics[width=0.8\columnwidth]{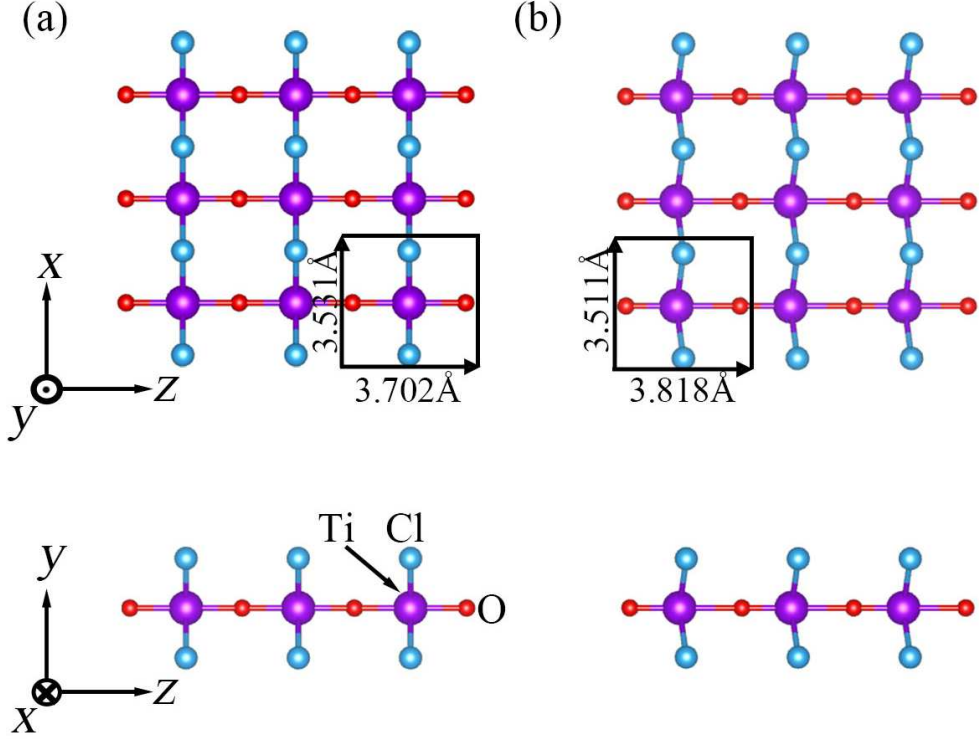}
\caption{\label{fig:fig1} Top and side views of monolayer TiOCl$_2$ in the (a) paraelectric and (b) ferroelectric state. Black rectangles denote the unit cell, with the optimized lattice constants marked.}
\end{figure}

\begin{figure}[htb]
	\includegraphics[width=0.8\columnwidth]{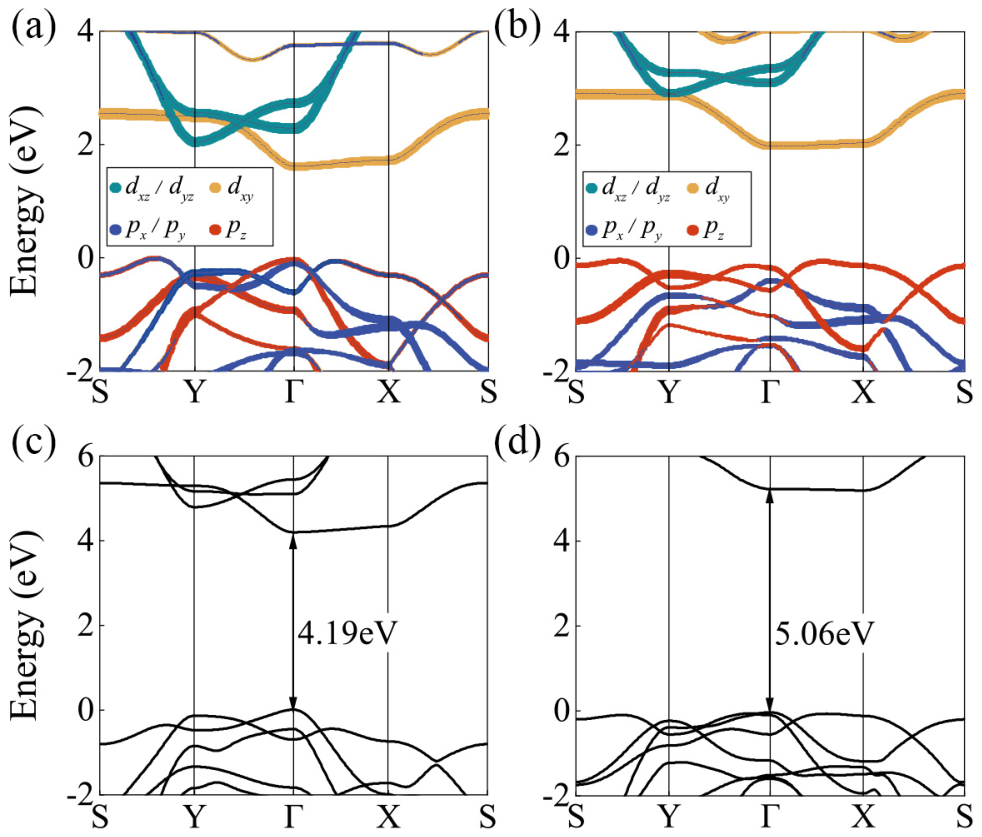}
	\caption{\label{fig:fig2} (Color online) (a)[(b)] The PBE fatband and (c)[(d)] $G_0$$W_0$ quasiparticle band structures of monolayer TiOCl$_2$ in the paraelectric [ferroelectric] state. The valence band top is set to energy zero.}
\end{figure}

Figures 1(a) and 1(b) show the geometries of monolayer TiOCl$_2$ in its paraelectric and ferroelectric state, respectively, both of which belong to the orthorhombic crystal system. The space group symmetry of the paraelectric structure is $Pmmm$ with the inversion center and optimized lattice constants of 3.702 and 3.531 \AA. The presence of imaginary frequencies in its phonon spectrum indicates that the structure is unstable\cite{Tan}. The ferroelectric structure shows a spontaneous displacement of Ti atoms along the O-Ti-O line [see Fig. 1(b)], and two different Ti-O bonds with lengths of 2.12 and 1.70 \AA, respectively, appear. Its space group symmetry becomes $Pmm2$ and no longer has an inversion center, with optimized lattice constants of 3.818 and 3.511 \AA. For convenience, we set the polarization direction to the $z$-axis and the in-plane vertical direction to $x$-axis (see Fig. 1). The ferroelectric polarization slightly stretches the lattice along $z$-axis by 3\%, while the lattice distortion along $x$-axis is an order of magnitude smaller. The total energy of the ferroelectric structure is reduced by 78 meV compared to the paraelectric one and there are no more imaginary frequencies in the phonon spectrum, indicative of the structural stability.

Figures 2(a) and 2(b) show the fatband structures of monolayer TiOCl$_2$ in the paraelectric and ferroelectric states, respectively, from which the orbital contributions of different energy states are visualized. The paraelectric band has a direct gap of 1.62 eV at the $\Gamma$ point, while the minimum gap of the ferroelectric band is indirect, with the conduction band bottom at the $\Gamma$ point and the valence band top at the point along the $SY$ line. Quantitatively, this indirect gap is only 20 meV smaller than the direct gap of 2.02 eV at the $\Gamma$ point.

Besides widening the gap, another obvious effect of spontaneous polarization is that it leads to the splitting of the highest energy state at the $\Gamma$ point. The valence band maximum of the paraelectric state is doubly degenerate, where two bands contributed mainly by O/Cl $p_z$ and $p_x$/$p_y$ orbitals intersect. Our later calculations show that this double degeneracy is accidental. Whilst, the two bands $p_z$ and $p_x$/$p_y$ in the ferroelectric state show an energy splitting of 222 meV at the $\Gamma$ point. Comparing the two bands in the paraelectric and ferroelectric states reveals that the $p_z$ states in the latter is overall closer to the Fermi energy compared to the former. As we will see later, this strongly affects the optical transition. On the other hand, it can be seen that both the paraelectric and ferroelectric states have their lowest conduction band of the Ti $d_{xy}$ feature with a dispersion of about 1 eV. Such a large $d$-band dispersion implies a strong hybridization between the Ti $d$ and the ligand O/Cl $p$ orbitals.

It is well known that conventional PBE calculations tend to underestimate the gaps of transition-metal compounds\cite{TanJCP,Bu}. For example, previous HSE study of ferroelectric monolayer TiOCl$_2$ yielded a larger minimum gap of 3.50 eV\cite{Tan}. To remedy the gap problem, we further calculate the $GW$ quasiparticle band structure as plotted in Figs. 2(c) and 2(d). One can see that the $\Gamma$-point direct gap reaches 4.19 and 5.06 eV for the paraelectric and ferroelectric states, which are increased by 2.57 and 3.04 eV, respectively, compared to those of the PBE. Now, there is no longer double degeneracy for the highest energy states at the $\Gamma$ point, even for the paraelectric case, unlike that of the PBE. Quantitatively, the energy splitting 454 meV of the paraelectric state is more than five times 82 meV of the ferroelectric state. This manifests itself as an orbit-dependent quasiparticle correction, which has been reported in other low-dimensional systems\cite{LiuJ,YangL,Prezzi,LiuJ22}. In this sense, the double degeneracy at the valence band top of the paraelectric case obtained by the PBE is indeed accidental. As will be seen later, a small strain can also eliminate this double degeneracy. Apart from these, the $GW$ band follows the same general trend as the PBE one.

\begin{figure*}[htp]
	\includegraphics[width=0.95\columnwidth]{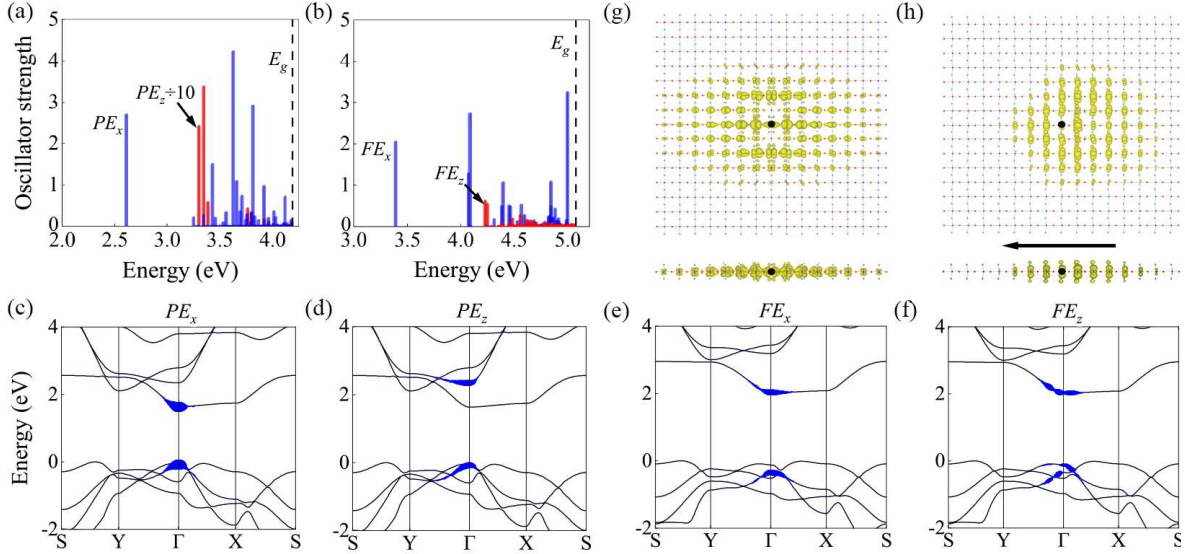}
	\caption{\label{fig:fig3} The calculated exciton energy spectrum inside the quasiparticle gap (Black vertical dashed lines) and the corresponding relative oscillator strength by solving the BSE under the incident light polarization along $x$- and $z$-axis for monolayer TiOCl$_2$, respectively, in the (a) paraelectric and (b) ferroelectric state. The BSE fatband structure for the (c) $PE_x$, (d) $PE_z$, (e) $FE_x$, and (f) $FE_z$ exciton. It is superimposed on the PBE one-electron band to visually inspect what kind of electron and hole states dominates the contribution to the very exciton. See the main text for details. The thicker the blue line, the higher the contribution. Real-space exciton wave functions modulus of (g) $PE_z$ and (h) $FE_z$ excitons, with the hole fixed at the center (the black dots). Black arrow denotes the direction of the built-in electric field in the ferroelectric state.}
\end{figure*}

Next, we solve the BSE on top of the quasiparticle band structures to obtain the excitonic properties. Since the effect of ferroelectricity is of interest here, we consider two special polarization cases when light is incident perpendicular to the basal plane, i.e., polarization parallel to (along the $z$-axis) and perpendicular to ferroelectric direction (along the $x$-axis). Figures 3(a) and 3(b) show the calculated exciton energy spectrum below the quasiparticle gap for the paraelectric and ferroelectric states, respectively.

The first absorption peaks of the paraelectric case appear at 2.63 and 3.35 eV, corresponding to $x$- and $z$-axis incidence, respectively (denoted as $PE_x$ and $PE_z$ hereafter). But the latter has an oscillator strength that is an order of magnitude higher. The first absorption peaks of the ferroelectric case are located at 3.40 and 4.25 eV under $x$- and $z$-axis incidence, respectively (denoted as $FE_x$ and $FE_z$ hereafter). In this case, the oscillator strength is at the same level for both. All the four peaks are inside the minimum direct quasiparticle gap (Black vertical dashed lines) and therefore correspond to exciton absorption.

Spontaneous displacement changes the Ti local symmetry from $D_{2 \rm h}$ to $C_{2 \rm v}$. An exhaustive group theory analysis tells that under both cases the optical transitions between $p_z$ and $d_{xy}$ states are forbidden while the optical transitions between $p_x$/$p_y$ and $d_{xy}$ states are allowed. For the sake of clarity, we show the BSE fatband structures in Figs. 3(c)-3(f) to visualize which \textbf{k}-point electron-hole pair dominates the contribution to the very exciton. Note that we solve the BSE on top of the $GW$ band but here the exciton fatband is superimposed to the PBE band, which facilitates a direct connection to the atomic orbital contributions as shown in Figs. 2(a) and 2(b). It is found that the $PE_x$ contains mainly the valence band top $p_x$/$p_y$ hole state with the conduction band bottom $d_{xy}$ electron state. However, the $PE_z$ is essentially different in that it consists mainly of valence band top $p_x$/$p_y$ hole state and the second lower conduction band $d_{xz}$/$d_{yz}$ electron state. The group theory analysis shows that the transitions between the two are indeed optically allowed.

In contrast, the electronic states of both $FE_x$ and $FE_z$ excitons come from $d_{xy}$ and their differences are all in the hole states. While the hole states both have a $p_x$/$p_y$ contribution, $FE_z$ also contains a part of the highest $p_z$ state along $\Gamma$$X$. This is interesting. On the one hand, this means that the difference in the one-electron gap corresponding to the $FE_x$ and $FE_z$ excitons is very small (only 82 meV for the splitting at the $\Gamma$ point in the $GW$ band structure). However, the difference in their transition energies is as high as 4.25 - 3.40 = 0.85 eV. Thus, the difference in binding strength between $FE_x$ and $FE_z$ excitons must be obvious. On the other hand, as mentioned before, the group theory analysis shows that the optical transition between $p_z$ and $d_{xy}$ is forbidden. This is indeed the case at the $\Gamma$ point. However, deviating from the $\Gamma$ point, especially close to where the $p_z$ band intersects the $p_x$/$p_y$ band, the hole consists mainly of the $p_z$ states. These two points suggest that spontaneous polarization not only significantly affects exciton binding, but also the selection rule.

\begin{table}[htbp]
	\caption{Exciton transition energy ($E_S$(0), in eV), corresponding one-electron gap ($E_g$, in eV), exciton binding energy ($E_b$, in eV), square of the dipole matrix element (${\mu_S^2}$, in a.u.), and radiative lifetimes ($\tau$, in ps) of the four characterized excitons induced by $x$- or $z$-axis incidence to the monolayer TiOCl$_2$ in its paraelectric and ferroelectric state.}
	\renewcommand\arraystretch{1.5}
	\setlength\tabcolsep{9pt}
	\begin{tabular}{ccccccccccc}
		\hline \hline
		Exciton &$E_S$(0) ~&$E_g$ ~&$E_b$ ~&${\mu_S^2}$ ~&$\tau$  \\
		\hline
		$PE_x$  &{2.63} &{4.65}  &{2.02} &{0.16} & {0.39}  \\
		$PE_z$  &{3.35} &{5.55}  &{2.20} &{1.46} &{0.03} \\
		$FE_x$  &{3.40} &{5.14}  &{1.74} &{0.13} &{0.40} \\
		$FE_z$  &{4.25} &{5.10}  &{0.85} &{0.03} &{1.22} \\		
		\hline
		\hline
	\end{tabular}
\end{table}

Exciton binding energy and lifetime are the two most important parameters characterizing the excitonic properties\cite{Spataru,Hsiao1,Hsiao2,Palummo1,Olsen,He,Zhou,Choi}. The exciton binding energy is defined as the difference between the corresponding one-electron energy gap and the transition energy\cite{Jiang2017}. The former can be inferred from the fatband, while the latter is a direct consequence of solving the BSE.  We list the binding energies of the four characteristic excitons in Table 1. We find them all around $\sim$2 eV, with the exception of $FE_z$, whose binding energy is only about half that, i.e., 0.85 eV. This is somewhat interesting because it happens to be the exciton in the direction of spontaneous polarization.

We also calculate their radiative lifetimes according to Eqs. (1) and (2) and the results are summarized in Table 1. One can see that the lifetimes of $PE_x$ and $FE_x$ excitons are similar at 0.39 and 0.40 ps, respectively. But for $z$-axis incidence, the lifetime of $PE_z$ exciton is only 0.03 ps while the one of $FE_z$ is as high as 1.22 ps. The latter is about 40 times higher than the former. It is interesting to note that the exciton lifetime also changes dramatically in the direction of spontaneous polarization, while in the vertical direction the exciton lifetime remains almost unchanged. To visually compare the differences between $PE_z$ and $FE_z$ excitons, we plot their real-space wavefunctions in Figs. 3(g) and 3(h), respectively. It is clear that for the centrosymmetric paraelectric state, the electrons are symmetrically distributed around the central hole. However, for the ferroelectric state, the electrons accumulate in the opposite direction of the built-in electric field generated by the spontaneous polarization. This is a direct manifestation of the effect of the built-in electric field on the excitonic state.

\begin{figure*}[htp]
	\includegraphics[width=0.95\columnwidth]{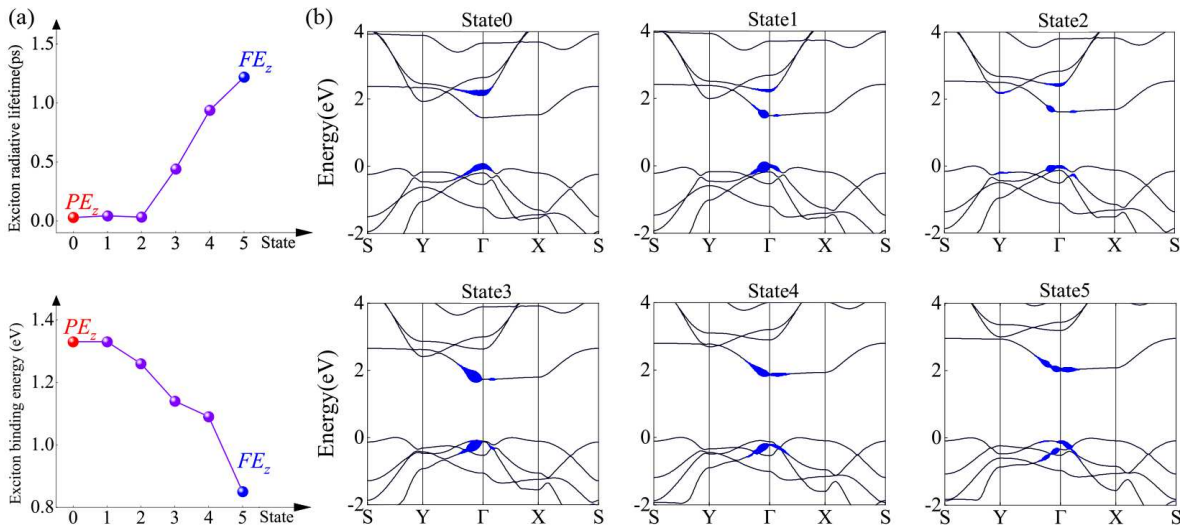}
	\caption{\label{fig:fig4} (a) Exciton lifetime (Top panel) and binding energy (Bottom panel) as a function of different state configurations. The first state ``0" and the last state ``5" represent the paraelectric and ferroelectric states, respectively. The lattice constants of the ferroelectric state are adopted for the paraelectric state and the four intermediate states. (b) Exciton fatband structures corresponding to the six states in (a). The conventions used are the same as in Figs. 3(c)-3(f).}
\end{figure*}

Counting all the physical constant terms as $K$, Eq. (1) can be rewritten as
\begin{equation}
	{\tau} = K\frac{{{A_{uc}}}}{{{E_S(0)}\mu_S^2}},
	\label{eq:2}
	~~~~ K = \frac{\hbar^2c}{8\pi e^2}~~.
\end{equation}
Clearly, the $\tau$ is determined by the three terms $A_{uc}$, $E_S$(0) and $\mu_S^2$. First, the variation of $A_{uc}$ produced by the different lattice constants of the paraelectric and ferroelectric states is only 3\%, which is negligible with respect to the 40-fold increase of the exciton lifetime. Second, as can be seen from Table 1, all transition energies $E_S$(0) are in the same order of magnitude, again not enough to cause a change of tens of times. There is no doubt that the giant enhancement of the exciton lifetime originates from $\mu_S^2$. We list in Table 1 the calculated $\mu_S^2$ for the different excitons. It is found that the $\mu_S^2$ of the $PE_z$ exciton is as high as 1.46 a.u., while that of the $FE_z$ exciton is as low as 0.03 a.u.. They are one order of magnitude higher and one order of magnitude lower than the other two $x$-axis incidence excitons, respectively. Since the $\tau$ is inversely proportional to $\mu_S^2$, a larger/smaller $\mu_S^2$ implies a smaller/larger lifetime. This is physically easy to understand. Equation (2) shows that the $\mu_S^2$ is proportional to the exciton transition dipole, which characterises the recombination probabilities of the electron-hole pairs that make up the exciton. A lower recombination rate naturally leads to a longer exciton lifetime.

To explore in depth the correlation between ferroelectricity and exciton binding energy reduction and lifetime enhancement, we manually set the displacement of Ti atoms along the O-Ti-O chain to obtain ferroelectric configurations with different polarization strengths, and then calculate their excitonic properties. In this way, we are able to monitor the dependence of excitonic properties on ferroelectric polarization. We have inserted four intermediate states uniformly between the paraelectric and ferroelectric states and calculated the light absorption spectrum under $z$-axis incidence for each configuration, from which we derive the exciton binding energy and lifetime, as summarized in Fig. 4(a). Note that the lattice constants of the paraelectric and ferroelectric states are slightly different. To complete the above insertion, the lattice constants of the ferroelectric state are adopted for all configurations. In consequence, the results for the paraelectric state in Fig. 4 are actually obtained under a small strain. Compared to the data we obtained earlier, it is obvious that this does change the electronic structure and excitonic properties of the paraelectric state. However, the trend of the effect of ferroelectricity on the exciton binding energy and lifetime, as well as the physics behind it, which is the focus of this work, has not been altered.

As can be seen in Fig. 4(a), the exciton lifetime does not increase immediately when the Ti atom is displaced, but remains approximately the same as the paraelectric state for the first two intermediate states. After that, it increases rapidly and is accompanied by a continuous decrease in exciton binding energy. Although the exciton lifetime and exciton binding energy are not explicitly related in Eq. (1), the general consensus is that higher exciton binding energies correspond to longer exciton lifetimes. Clearly, our findings here are different from the usual understanding. In order to figure out the physical reason behind this, we plot the excitonic fatband for each state in Fig. 4(b).

Firstly, let us focus our attention on the effect of lattice strain on the paraelectric state. Compared to Fig. 2(a), the strain removes the two-fold degeneracy of the valence band top, causing an energy splitting of 248 meV at the $\Gamma$ point, similar to the $GW$ result [see Fig. 2(c)]. The band structure still exhibits a $\Gamma$-point direct gap, with the value slightly changed from 1.62 eV to 1.66 eV. The quasiparticle correction gives the gap of 3.29 eV which is narrowed by 0.9 eV compared to the strain-free case. Solving the BSE yields transition energies of 2.10 and 2.85 eV for $PE_x$ and $PE_z$ excitons, respectively, with the corresponding binding energies of 1.18 and 1.33 eV. The $PE_x$ exciton lifetime is now 0.63 ps, which is not significantly different from the 0.39 ps in the absence of strain. And for $PE_z$ the exciton lifetime is almost the same in both cases. These results suggest that strain does affect the electronic structure and even changes the exciton binding energy dramatically, but the effect on the exciton lifetime is limited.

The fatband evolution in Fig. 4(b) unambiguously reflects the significant change in exciton characteristics during the transition from the paraelectric to the ferroelectric state. The $PE_z$ of the paraelectric state consists of the second lowest $d_{xz}$/$d_{yz}$ electron and the highest $p_x$/$p_y$ hole. The situation changes as the Ti displacement increases, and at the third intermediate state (State3), the electron and hole characteristics appear to change substantially. The sudden increase in exciton lifetime occurs at precisely this point. More specifically, for the electron state, the $d_{xz}$/$d_{yz}$ component becomes less and less, while the lowest energy $d_{xy}$ component gradually increases. After State3, the former disappears completely, while the electron state is entirely contributed by $d_{xy}$. For the hole, it is initially dominated entirely by the highest $p_x$/$p_y$ state. However, from State0 to State3, the splitting of $p_x$/$p_y$ and $p_z$ bands decreases until they intersect and then exchange order, implying a strong hybridization between them. Thereafter, the hole becomes a mixture of $p_x/p_y$ and $p_z$ states.

The above analysis of the electron/hole characteristics shows that the exciton undergoes a sudden change in nature at State3. In other words, it can be considered that there the $PE_z$ exciton is destroyed and the $FE_z$ exciton is nascent. Generally, spontaneous polarization has two consequences. One is the built-in electric field, which always tends to perturb the exciton by inhibiting electron-hole binding. When this built-in field reaches a critical value, the $PE_z$ exciton will be completely broken. Another is structural distortion, which reduces the symmetry of the system while enhancing orbital hybridization, so that the exciton consisting of $p_z$ hole and $d_{xy}$ electron, which otherwise does not satisfy the optical selection rules, acquires optical activity after the appearance of a $p_x$/$p_y$ component in the hole. This structural distortion leads to a relaxation of the selection rule and excites excitons that are otherwise not optically active. In principle, the $\mu_S^2$ of such excitons is smaller, giving rise to longer exciton lifetimes. Thus, it is the synergistic effect of the built-in electric field and the structural distortion that leads to the destruction-regeneration of the exciton, which together leads to a reduction in binding energy and an increase in lifetime.

In summary, we have investigated the effect of ferroelectric polarization on exciton lifetime and exciton binding energy with a prototype monolayer TiOCl$_2$ using first-principles calculations. We find that this effect depends on the direction of the incident light polarization relative to the ferroelectric polarization. When the two are perpendicular, both effects are very small. However, when the two are parallel, spontaneous ferroelectric polarization leads to a 40-fold increase in exciton lifetime and a reduction in exciton binding energy to half. The effect of ferroelectric polarization is manifested in two ways. On the one hand, it introduces a built-in electric field that tends to dissociate the exciton and reduce the binding energy. On the other hand, structural distortions lower the system symmetry and enhance orbital hybridization, which relaxes the selection rule and leads to the emergence of new long-lived excitons. Our study not only provides new insights into the photovoltaic properties of ferroelectric materials, but also has important implications for the use of ferroelectricity to modulate excitonic properties.

\begin{acknowledgments}
We thank Shudong Wang for valuable discussions. This work was supported by the Ministry of Science and Technology of China (Grant No. 2020YFA0308800) and the National Natural Science Foundation of China (Grant No. 12074034).
\end{acknowledgments}


\begin{thebibliography}{90}%
\makeatletter
%{\bibliographystyle{plain}
	%\bibliography{1}
\bibitem{Klimov} V. I. Klimov, S. A. Ivanov, J. Nanda, M. Achermann, I. Bezel, J. A. McGuire, and A. Piryatinski, Nature \textbf{447}, 441 (2007).

\bibitem{Sambur} J. B. Sambur, T. Novet, and B. A. Parkinson, Science \textbf{330}, 63 (2010).

\bibitem{EI1967} D. J\'{e}rome, T. M. Rice, and W. Kohn, Phys. Rev. \textbf{158}, 462 (1967).

\bibitem{Kogar} A. Kogar, S. Vig, M. S. Rak, A. A. Husain, F. Flicker, Y. I. Joe, L. Venema, G. J. MacDougall, T. C. Chiang, E. Fradkin, J. van Wezel, and P. Abbamonte, Science \textbf{358}, 1314 (2017).

\bibitem{Lee} M. M. Lee, J. Teuscher, T. Miyasaka, T. N. Murakami, and H. J. Snaith, Science \textbf{338}, 643 (2012).

\bibitem{Jarvist} J. M. Frost, K. T. Butler, F. Brivio, C. H. Hendon, M. van Schilfgaarde, and A. Walsh, Nano Lett. \textbf{14}, 2584 (2014).

\bibitem{Noel} N. C. Giebink, G. P. Wiederrecht, M. R. Wasielewski, and S. R. Forrest, Phys. Rev. B \textbf{83}, 195326 (2011).

\bibitem{Mak} K. F. Mak, C. Lee, J. Hone, J. Shan, and T. F. Heinz, Phys. Rev. Lett. \textbf{105}, 136805 (2010).

\bibitem{Jiang2017} Z. Y. Jiang, Z. R. Liu, Y. C. Li, and W. H. Duan, Phys. Rev. Lett. \textbf{118}, 266401 (2017).

\bibitem{Jiang2018} Z. Y. Jiang, Y. C. Li, S. B. Zhang, and W. H. Duan, Phys. Rev. B \textbf{98}, 081408(R) (2018).

\bibitem{Jiang2019} Z. Y. Jiang, Y. C. Li, W. H. Duan, and S. B. Zhang, Phys. Rev. Lett. \textbf{122}, 236402 (2019).

\bibitem{Jiang2020} Z. Y. Jiang, W. K. Lou, Y. Liu, Y. C. Li, H. F. Song, K. Chang, W. H. Duan, and S. B. Zhang, Phys. Rev. Lett. \textbf{124}, 166401 (2020).

\bibitem{Varsano}  D. Varsano, M. Palummo, E. Molinari, and M. Rontani, Nature Nanotechnol. \textbf{15}, 367 (2020).

\bibitem{Dong2020} S. Dong and Y. C. Li, Phys. Rev. B \textbf{102}, 155119 (2020).

\bibitem{Dong2021} S. Dong and Y. C. Li, Phys. Rev. B \textbf{104}, 085133 (2021).

\bibitem{Chang} K. Chang, J. W. Liu, H. C. Lin, N. Wang, K. Zhao, A. M. Zhang, F. Jin, Y. Zhong, X. P. Hu, W. H. Duan, Q. M. Zhang, L. Fu, Q.-K. Xue, X. Chen, and S.-H. Ji, Science \textbf{353}, 274 (2016).

\bibitem{Fei} Z. Fei, W. Zhao, T. A. Palomaki, B. Sun, M. K. Miller, Z. Zhao, J. Yan, X. Xu, and D. H. Cobden, Nature \textbf{560}, 336 (2018).

\bibitem{Ponomarev} I. V. Ponomarev, L. I. Deych, and A. A. Lisyansky, Phys. Rev. B \textbf{72}, 115304 (2005).

\bibitem{Chernikov} A. Chernikov, A. M. van der Zande, H. M. Hill, A. F. Rigosi, A. Velauthapillai, J. Hone, and T. F. Heinz, Phys. Rev. Lett. \textbf{115}, 126802 (2015).

\bibitem{Raja} A. Raja, A. Chaves, J. Yu, G. Arefe, H. M. Hill, A. F. Rigosi, T. C. Berkelbach, P. Nagler, C. Sch\"{u}ller, T. Korn, C. Nuckolls, J. Hone, L. E. Brus, T. F. Heinz, D. R. Reichman, and A. Chernikov, Nat. Commun. \textbf{8}, 15251 (2017).

\bibitem{Massicotte} M. Massicotte, F. Vialla, P. Schmidt, M. B. Lundeberg, S. Latini, S. Haastrup, M. Danovich, D. Davydovskaya, K. Watanabe, T. Taniguchi, V. I. Fal$^\prime$ko, K. S. Thygesen, T. G. Pedersen, and F. H. L. Koppens, Nat. Commun. \textbf{9}, 1633 (2018).

\bibitem{LiuJ} J. Liu, G.-B. Liu, and Y. C. Li, Phys. Rev. B \textbf{104}, 085150  (2021).

\bibitem{Yang} S. Y. Yang, J. Seidel, S. J. Byrnes, P. Shafer, C.-H. Yang, M. D. Rossell, P. Yu, Y.-H. Chu, J. F. Scott, J. W. Ager, III, L. W. Martin, and R. Ramesh, Nature Nanotech. \textbf{5}, 143 (2010).

\bibitem{Zenkevich} A. Zenkevich, Yu. Matveyev, K. Maksimova, R. Gaynutdinov, A. Tolstikhina, and V. Fridkin, Phys. Rev. B \textbf{90}, 161409(R) (2021).

\bibitem{Seidel} J. Seidel, D. Y. Fu, S.-Y. Yang, E. Alarc\'{o}n-Llad\'{o}, J. Q. Wu, R. Ramesh, and J. W. Ager III, Phys. Rev. Lett. \textbf{107}, 126805 (2011).

\bibitem{Bhatnagar} A. Bhatnagar, A. R. Chaudhuri, Y. H. Kim, D. Hesse, and M. Alexe, Nat. Commun. \textbf{4}, 2835 (2013).

\bibitem{PBE} J. P. Perdew, K. Burke, and M. Ernzerhof, Phys. Rev. Lett. \textbf{77}, 3865 (1996).

\bibitem{QE} P. Giannozzi, S. Baroni, N. Bonini, M. Calandra, R. Car, C. Cavazzoni, D. Ceresoli, G. L. Chiarotti, M. Cococcioni, I. Dabo, A. D. Corso, S. de Gironcoli, S. Fabris, G. Fratesi, R. Gebauer, U. Gerstmann, C. Gougoussis, A. Kokalj, M. Lazzeri, L. Martin-Samos et al., J. Phys.: Condens. Matter \textbf{21}, 395502 (2009).

\bibitem{GWBSE} M. Rohlfing and S. G. Louie, Phys. Rev. B \textbf{62}, 4927 (2000).

\bibitem{Yambo1} A. Marini, C. Hogan, M. Gr\"{u}ning, and D. Varsano, Comput. Phys. Commun. \textbf{180}, 1391 (2009).

\bibitem{Yambo2}  D. Sangalli, A. Ferretti, H. Miranda, C. Attaccalite, I. Marri, E. Cannuccia, P. Melo, M. Marsili, F. Paleari, A. Marrazzo, et al., J. Phys.: Condens. Matter \textbf{31}, 325902 (2019).

\bibitem{Palummo1} M. Palummo, M. Bernardi, and J. C. Grossman, Nano Lett. \textbf{15}, 2794 (2015).

\bibitem{Hsiao1} H. Y. Chen, M. Palummo, D. Sangalli, and M. Bernardi, Nano Lett. \textbf{18}, 3839 (2018).

\bibitem{Hsiao2} H. Y. Chen, V. A. Jhalani, M. Palummo, and M. Bernardi, Phys. Rev. B \textbf{100}, 075135 (2019).

\bibitem{Spataru} C. Spataru, S. Ismail-Beigi, R. Capaz, and S. G. Louie, Phys. Rev. Lett. \textbf{95}, 247402 (2005).

\bibitem{Tan} H. X. Tan, M. L. Li, H. T. Liu, Z. R. Liu, Y. C. Li, and W. H. Duan, Phys. Rev. B \textbf{99}, 195434 (2019).

\bibitem{TanJCP}H. Tan, H. Liu, Y. Li, W. Duan, and S. Zhang, J. Chem. Phys. \textbf{151}, 124703 (2019).

\bibitem{Bu} X. T. Bu and Y. C. Li, Phys. Rev. B \textbf{106}, L241101 (2022).

\bibitem{YangL} L. Yang, M. L. Cohen, and S. G. Louie, Nano Lett. \textbf{7}, 3112 (2007).

\bibitem{Prezzi} D. Prezzi, D. Varsano, A. Ruini, A. Marini, and E. Molinari, Phys. Rev. B \textbf{77}, 041404(R) (2008).

\bibitem{LiuJ22} J. Liu and Y. C. Li, Phys. Rev. B \textbf{106}, 035135  (2022).

\bibitem{He} K. He, N. Kumar, L. Zhao, Z. Wang, K. F. Mak,  H. Zhao and J. Shan, Phys. Rev. Lett. \textbf{113}, 026803  (2014).

\bibitem{Olsen} T. Olsen, S. Latini, F. Rasmussen, and K. S. Thygesen, Phys. Rev. Lett. \textbf{116}, 056401 (2016).

\bibitem{Zhou} Y. Zhou, G. Scuri, J. Sung, R. Gelly, D. Wild, K. De Greve, A. Joe, T. Taniguchi, K. Watanabe, P. Kim, M. Lukin, and H. Park, Phys. Rev. Lett. \textbf{124}, 027401 (2020).

\bibitem{Choi} J. Choi, M. Florian, A. Steinhoff, D. Erben, K. Tran, D. S. Kim, L. Sun, J. Quan, R. Claassen, S. Majumder, J. A. Hollingsworth, T. Taniguchi, K. Watanabe, K. Ueno, A. Singh, G. Moody, F. Jahnke, and X. Q. Li, Phys. Rev. Lett. \textbf{126}, 047401 (2021).



\end{thebibliography}
\end{document}